\documentclass{article}
\usepackage{chicago}
\usepackage{amssymb}

\newcommand{\iec}{\mbox{i.\,e.\,}}

\newcommand{\egc}{\mbox{e.\,g.\,}}

\newcommand{\dr}[1]{\ensuremath{\mathrm{d} #1\,}}

\newcommand{\dbd}[2]{\ensuremath{\frac{\dr{#1}}{\dr{#2}}}}

\newcommand{\pbp}[2]{\ensuremath{\frac{\partial #1}{\partial #2}}}

\newcommand{\ket}[1]{\ensuremath{\left|  #1 \right\rangle}}
\newcommand{\bra}[1]{\ensuremath{\left\langle #1 \right|}}

\newcommand{\matel}[3]{\ensuremath{\bra{#1} #2 \ket{#3}}}
 
\newcommand{\op}[1]{\ensuremath{\widehat{\textsf{\ensuremath{#1}}}}}

\newcommand{\be}{\begin{equation}}
\newcommand{\ee}{\end{equation}}
\newcommand{\e}[1]{\mathrm{e}^{#1}}

\begin{document}
\title{The case for black hole thermodynamics \\  Part I: phenomenological thermodynamics}
\author{David Wallace\thanks{Dornsife College of Letters, Arts and Sciences, University of Southern California; email \texttt{dmwallac@usc.edu}}}
\maketitle

\begin{abstract}
I give a fairly systematic and thorough presentation of the case for regarding black holes as thermodynamic systems in the fullest sense, aimed at readers with some familiarity with thermodynamics, quantum mechanics and general relativity but not presuming advanced knowledge of quantum gravity. I pay particular attention to (i) the availability in classical black hole thermodynamics of a well-defined notion of adiabatic intervention; (ii) the power of the membrane paradigm to make black hole thermodynamics precise and to extend it to local-equilibrium contexts; (iii) the central role of Hawking radiation in permitting black holes to be in thermal contact with one another; (iv) the wide range of routes by which  Hawking radiation can be derived and its back-reaction on the black hole calculated; (v) the interpretation of Hawking radiation close to the black hole as a gravitationally bound thermal atmosphere. In an appendix I discuss recent criticisms of black hole thermodynamics by Dougherty and Callender. This paper confines its attention to the thermodynamics of black holes; a sequel will consider their statistical mechanics. 
\end{abstract}

\section{Introduction}

Black hole thermodynamics (BHT) is perhaps the most striking and unexpected development in the theoretical physics of the last forty years. It combines the three main areas of `fundamental' theoretical physics --- quantum theory, general relativity, and thermal physics --- and it offers a conceptual testing ground for quantum gravity that might be the nearest that field has to experimental evidence. Yet BHT itself relies almost entirely on theoretical arguments, and its most celebrated result --- Hawking's argument that black holes emit radiation --- has no direct empirical support and little prospect of getting any. So to outsiders --- to physicists in other disciplines, or to philosophers of science --- the community's confidence in BHT can seem surprising, or even suspicious. Can we really be so confident of anything without any grounding in observation?

In this article, and its sequel, I want to lay out as carefully and thoroughly as I can the theoretical evidence for BHT. It is written with the zeal of the convert: I began this project sharing at least some of the outsiders' scepticism, and became persuaded that the evidence is enormously strong both that black holes are thermodynamical systems in the fullest sense of the word, and that their thermodynamic behaviour has a statistical-mechanical underpinning in quantum gravity (and, as a consequence, that black hole evaporation is a unitary process not different in kind from the cooling of other hot systems, and that it involves no fundamental loss of information). 

There are of course many reviews of this material. But those I know either (i) take for granted the main results of BHT, moving quickly over established material to get students up to speed with the research frontier; (ii) are explicitly historical, which illuminates how the community \emph{in fact} came to accept BHT but can obscure the logic of whether and why they \emph{should have} accepted it, or (iii) are written at a very high level of mathematical rigor, so high that a large fraction of the literature has to be omitted. I hope this paper will be complementary to extant material. With few exceptions, I present and describe results without going into the details of their derivation, and the student who wishes to properly understand the topic will need to read this paper in parallel with some of the extant review literature. My starting points (for this part of the paper) were \citeN{harlowreview}, Jacobson~\citeyear{jacobsonblackholelectures,jacobsonqft}, \citeN{membraneparadigm}, and Wald~\citeyear{waldqft,wald2001}.

A note on mathematical rigor: the tendency in foundational work on this subject (see, \egc, \citeN{belotetalinformationloss} and \citeN{earmanunruh}) has been to work at the level of rigor typical in mathematical physics, where all results are stated exactly and proved rigorously. This is much higher than the standard in theoretical physics more generally; it has the advantage of reliability, but the disadvantage that a very large fraction of the literature must be elided --- especially in a frontier area like this, where the underlying physical principles are unclear and the mathematical framework partial and under active development. And the case for BHT --- as will become apparent throughout this paper and, even more so, its sequel --- rests not so much on individual results that have been established with full precision and rigor, but on the many independent calculations with different premises and approximation schemes that all lead to the same result. So this paper is written at the theoretical-physics level; I hope that readers who prefer their mathematics more precise will at least get a sense as to why \emph{the community} takes BHT so seriously, even if they are not persuaded themselves.

This is a large topic, too large for any one paper. In this paper I confine my attention to phenomenological thermodynamics, setting aside any considerations of statistical-mechanical underpinnings for that thermodynamics. In \citeN{wallaceblackholestatmech} I consider the progress made in calculating the thermodynamical properties of black holes via statistical mechanics (in effective-field theory quantum gravity, in string theory, and via the AdS/CFT correspondence). And in \citeN{wallaceinformationloss} I use these two papers as a starting point to review and assess the notorious \emph{information-loss paradox} which has motivated a large part of the critical attention paid to BHT.

The structure of the paper is as follows. I begin in section \ref{review} by briefly reviewing classical thermodynamics, and discussing how it is modified for self-gravitating systems: to see whether black holes are thermodynamical, we need to be clear what thermodynamics is in the first place. In section \ref{classical} I consider classical black hole thermodynamics, arguing that while black holes offer a strikingly good realisation of the principles of thermodynamics when regarded as isolated systems, they completely fail to do so when considered as components of a larger system. In section \ref{quantum} I show how including the implications of quantum field theory, in particular (though not exclusively) the Hawking effect, entirely remove this limitation; I also review the strength of the evidence for the Hawking effect itself, and the related but logically stronger claim that Hawking radiation leads to black hole evaporation. In an appendix, I address the arguments of a recent paper by \citeN{doughertycallender} which criticises BHT (that paper was one trigger for my writing this paper, but engaging with its arguments in the main text would complicate my structure unhelpfully).

Readers familiar with extant debates on black hole thermodynamics may be surprised that in this paper I make virtually no mention of Bekenstein's classic argument \cite{Bekenstein1973} for black hole entropy on the grounds of information. Partly this is because this paper is confined to phenomenological thermodynamics, and the relation of information to thermodynamics is normally made at the statistical-mechanical level. But mainly it is just because the link between information and thermodynamics is \emph{controversial}, and so any argument for black hole thermodynamics from considerations of information is apt to inherit that controversy (recent critical takes on black hole thermodynamics by \citeN{wuthrichblackhole} and Dougherty and Callender (\emph{ibid}) both rely in one way or another on skepticism about the entropy-information link).Since (I will argue) we can make a compelling case for BHT and (in the sequel) black hole statistical mechanics without ever considering information, it seems simpler to sidestep the controversy. I discuss this in slightly more detail in the appendix.

I assume some familiarity with classical general relativity (in particular the Schwarzschild solution) and classical thermodynamics (and I quote standard results from both fields without explicit references); a little prior exposure to quantum field theory would also be helpful in section \ref{quantum}. Except where explicitly noted, I adopt units where $G=\hbar=c=k_B=1$.

\section{Thermodynamics and statistical mechanics: a brief review}\label{review}

Without any pretension to historical accuracy, complete precision or logical independence, we can break the salient parts of equilibrium thermodynamics into three: equilibrium and equilibration; the First and Second Laws for individual systems; interactions between multiple systems. (I believe the account I give basically tracks the consensus in mainstream physics; it broadly follows Wallace~\citeyear{wallacethermocontrol,wallaceactualstatmech}.) I discuss each in turn; I then briefly consider the generalisation of equilibrium thermodynamics to \emph{local} thermal equilibrium, and the subtleties introduced by gravitation. For this paper I do not need, and do not discuss, the statistical-mechanical underpinnings of thermodynamics.

\subsection{Equilibrium and equilibration}

 A thermodynamic system has a family of \emph{equilibrium states} parametrised by the energy and by a (usually small) number of additional conserved quantities and/or external constraints. In the absence of external interventions, if the system is in the equilibrium state corresponding to its constraints and conserved quantities, it remains in that state; if it is not, it \emph{equilibrates}, evolving towards that state and reaching it, to any given degree of accuracy, after a finite time (\citeN{BrownUffink2001} refer to this equilibration principle as the \emph{Minus First Law of Thermodynamics}).

 For instance, for a box of gas (of some fixed kind of particle) the external constraint is the volume of the box, and the conserved quantities are the energy, the number of particles, and in principle the momentum and angular momentum. In general we assume a nonrotating box and study it in its rest frame, and/or assume that the box is so massive not to be affected by particle collisions, so that momentum and angular momentum may be neglected and `energy' and `internal energy' can be identified; often we also take the particle number as fixed and do not include it explicitly as a variable.

\subsection{The First and Second Laws for individual systems}\label{firstsecondlaw}

Given an isolated thermodynamic system, an \emph{adiabatic} transformation of that system is some operation performed on the system, starting at equilibrium, that transforms its state to another equilibrium state without coupling it nontrivially to other thermodynamic systems.\footnote{Many presentations introduce the notion of ``heat'' as a primitive, and define adiabatic processes as those that do not involve heat transfer; I do not do so here because it is convenient for BHT to treat heat as a derived quantity.} Any such transformation can be thought of as a change to the external constraints and conserved quantities of the system via some external force; paradigm examples include expanding or compressing  a gas, or putting a non-rotating system into rotation.  The \emph{work done} by such a process is defined as the change in the system's energy, and (by conservation of energy) is then equal to the energy cost to the external agent.

Only some such changes are physically possible by means of adiabatic transformations. Specifically, if the system's equilibrium states are parameterised by energy $U$ and conserved quantities/external constraints $X_i$, there exists a function $S(U,X_1,\ldots X_N)$, called the \emph{entropy} of the system (and hence defined, as far as thermodynamics is concerned, \emph{only} at equilibrium), such that $S$ is non-decreasing under any adiabatic transformation. This entropy non-decrease law is one form of the \emph{Second Law of Thermodynamics}. 

Adiabatic transformations can then be broken into three categories: \emph{reversible} transformations, which leave $S$ unchanged; \emph{irreversible} transformations, which increase $S$, and \emph{thermodynamically forbidden transformations}, which decrease $S$. It is generally the case that all reversible and irreversible transformations are physically performable (at least in principle, and perhaps in an idealised limiting case) so that the Second Law imposes a necessary and sufficient condition for a transformation to be possible. In particular, if we make a very small adiabatic change to the $X_i$ and then wait for the system to re-equilibrate, that change will leave $S$ unchanged to a very high degree of accuracy. So sufficiently slow adiabatic changes to the $X_i$ will define processes which are very close to being reversible, becoming exactly reversible in the infinite-time limit. It is generally the case that such \emph{quasi-static} transformations are always available.

We can express the entropy in differential form as
\be
\mathrm{d}S = \beta\left(\mathrm{d}U + \sum_i \lambda_i X_i \right)
\ee
or, rearranging so that $U$ is a function of $S$ and the $X_i$,
\be\label{firstlaw}
\mathrm{d}U = T \mathrm{d}S - \sum_i \lambda_i X_i
\ee
where $T=1/\beta$. $T$ is called the \emph{thermodynamic temperature} and the $\lambda_i$ are the thermodynamic variables \emph{conjugate} to the $X_i$; they can be given explicitly by
\be
\frac{1}{T} = \left(\pbp{S}{U}\right)_{X_i}\,\,\,\, \lambda_i = T \left(\pbp{S}{X_i}\right)_{U,X_j}
\ee
or by
\be
T = \left(\pbp{U}{S}\right)_{X_i}\,\,\,\, \lambda_i = \left(\pbp{U}{X_i}\right)_{S,X_j}.
\ee
The $\lambda_i$ usually have a physical meaning: in particular, the variables conjugate to volume, momentum, angular momentum, particle number, and charge are, respectively, pressure, centre-of-mass velocity, angular velocity, chemical potential, and electric potential.

Equation (\ref{firstlaw}) is one form of the \emph{First Law of Thermodynamics}. It can be understood entirely statically, as a statement of the relations between different equilibrium states. But given the existence of quasi-static processes, we can also interpret it as describing the actual change in $U$ induced by small adiabatic changes $X_i \rightarrow X_i + \delta X_i$ to the parameters, together with a flow of energy $Q = T \delta S$ into the system from some external reservoir. Following \citeN[p.141]{waldqft}) we can call these the \emph{equilibrium-state} and \emph{physical-process} interpretations, respectively. Flow of energy of this kind is called \emph{heat flow} and makes sense even if the flow is not infinitesimal; conservation of energy entails that the change in a system's energy equals the heat flow into it plus the work done on it, which is another form of the First Law.

Finally, note that at this stage of our analysis $S$ (and, hence, $T$) is fixed only up to an arbitrary rescaling: we can replace $S$ with $f(S)$, for any smoothly increasing function $f$, and $1/T$ with $f'(1/T)$, without affecting anything said so far.

\subsection{Multiple thermodynamic systems}

Much of the content of thermodynamics is only available once we allow dynamical interactions between multiple systems. The rules for doing so are:
\begin{enumerate}
\item Any two systems may be placed in \emph{thermal contact}, so that heat may flow between them while their other conserved quantities and external parameters remain \emph{separately fixed}. This can be generalised to allow for other kinds of contact in which the two systems can exchange other conserved quantities.
\item Multiple systems in (perhaps-generalised) thermal contact may be treated as a single system; in particular, any such combined system will have a unique equilibrium state.
\item The Second Law of Thermodynamics generalises to require that the total entropy of two systems in (perhaps-generalised) thermal contact does not decrease when those systems exchange energy and other conserved quantities. For this to be well-defined, the possibility for rescaling of entropy decreases sharply: in multiple-system contexts, entropy must be taken as fixed up to a system-independent scale and a system-dependent additive constant. 
\end{enumerate}
From (2) and (3) together, it follows that:
\begin{enumerate}
\item[4.] If two systems are in thermal contact, and heat $\delta Q$ flows from system 1 to system 2, the total change in entropy is $\delta S = \delta Q (1/T_2 - 1/T_1)$. So heat will flow only if $T_1>T_2$, and indeed, no process can as its sole effect induce heat flow unless this condition holds (the \emph{Clausius statement} of the Second Law). It follows that a necessary and sufficient condition for two systems in thermal contact to be jointly at equilibrium is that they are separately at equilibrium with equal temperatures. (This generalises to other forms of contact.) As a consequence, the relation `at equilibrium with' is an equivalence relation: this is the \emph{Zeroth Law of thermodynamics}, and in textbook presentations is often taken as a starting point; in my presentation, it is a consequence of other assumptions.
\item [5.] Given a process involving an infinitesimal heat flow between two equilibrium systems at thermodynamic temperatures $T_1$, $T_2$ together with work $W$ done on the combined system, and such that the conserved quantities and external constraints of the two systems (other than energy) are unchanged at the end of the process, the First Law entails that
\be
W = T_1 \Delta S_1 + T_2 \Delta S_2 = T_1 \left( \Delta S_1 + \frac{T_2}{T_1}\Delta S_2\right).
\ee
Since the Second Law entails that $\Delta S_2 \geq - \Delta S_1$, we have
\be
W \geq T_1 \Delta S_1 \left (1 - (T_2/T_1)\right).
\ee
From this, we can read off that the maximum efficiency of any cyclical process which generates work from heat flow between the two systems is $(1-T_2/T_1)$ and, \emph{a fortiori}, that no cyclical process can as its sole effect convert heat flow from an equilibrium system into work done, which is the \emph{Kelvin statement} of the Second Law. (Other processes can do better, but they do not leave the other conserved quantities and constraints unchanged and so cannot be performed in a cycle.)
\end{enumerate}

\subsection{Local thermodynamic equilibrium}

In an extended body (such as a solid, a fluid, or a field), if the rate at which a small region of the fluid equilibrates is fast compared to the rate at which it exchanges energy and other conserved quantities with neighboring regions, the body will approach \emph{local thermal equilibrium}, at which we may express thermodynamic quantities like charge, energy, entropy, temperature and pressure as functions of position in the body. (For instance the sun, while not at equilibrium, is at local equilibrium, so that we can describe how temperature, pressure, entropy density and energy density vary from the core to the atmosphere.) Various phenomenological equations can be derived or postulated to describe the flow of thermodynamic quantities through the system. For instance, \emph{Ohm's Law} describes how current flow in a conductor is dissipated as heat, and the \emph{Navier-Stokes equations} describe the flow of a viscous fluid and the dissipation of organised energy as heat in that fluid. Various \emph{transport coefficients}, like electrical resistivity and viscocity, appear in those equations, so that they cannot simply be derived from the equation of state but require additional empirical input.

\subsection{Complications of gravity}\label{td-gravity}

Insofar as thermodynamics is the study of systems \emph{at equilibrium}, it has fairly few real-world applications (except black holes themselves) to systems in which gravity is the dominant interaction. Indeed, a well-known result in celestial mechanics (the \emph{gravothermal catastrophe} (\citeN{lyndenbellwood1968}; \citeN[pp.500-5]{binneygalacticdynamics} is a good introduction) demonstrates that classical Newtonian systems in which gravity is the only relevant form cannot reach stable equilibrium unless confined to a sufficiently small box. The only gravity-dominated astrophysical systems at thermal equilibrium (other than black holes themselves!) are degenerate-matter objects like white dwarfs and neutron stars, where quantum effects permit stability. Ordinary stars, for instance, are not at thermal equilibrium: there is a constant flow of energy from the core to the surface, and from the surface to interstellar space through emitted radiation; only the presence of fusion reactions in the core to replenish that lost heat allows stars to remain stable, until their fusion fuel is exhausted. 

Nonetheless, thermodynamics can be coherently formulated for the artificial, but well-defined, example of (relativistic or Newtonian) self-gravitating systems confined to a box, where the existence of long-range forces in these systems leads to important subtleties, even before we consider black holes. Rather than discuss the (somewhat controversial) general structure of these subtleties (for that discussion, see \citeN{wallacegravent}, \citeN{callenderheavy}, and references therein), I will illustrate them with a concrete example due to \citeN{sorkinwaldjiu}: a spherical box of radiation at thermal equilibrium, potentially large enough that self-gravitation has a discernible effect. 

The sphere is assumed to be nonrotating and at rest. Its equation of state depends on two parameters: its radius $R$ and its mass $M$ (for a relativistic system, and in units where 
$c=1$, its mass and its energy can be identified). A crucial parameter is the \emph{Schwarzschild radius} $R_S(M)=2GM/R$: if $R<R_S(M)$ then an event horizon forms around the sphere and it must be treated as a black hole. 

As long as $R \gg R_S(M)$, gravitating effects are fairly insignificant and the sphere may be treated as if it were non-self-gravitating. It then behaves as a pretty conventional thermodynamic system, with an extensive equation of state determined by the intensive formulae
\be\label{EMstate}
\rho=bT^4;\,\,\,\, s=\frac{4}{3}b T^3
\ee
that determine the energy density $\rho$ and entropy density $s$ as functions of the temperature. In particular, if we consider a sequence of successively larger spheres with $M/R^3$ held constant, the temperature and density of each sphere likewise remain constant. But for denser spheres (the transition occurs roughly around $R\simeq 5 R_S$) gravitational effects become highly important and the system displays several distinctive features characteristic of strongly self-gravitating systems (all discussed, or readily derived, in Sorkin \emph{et al}'s paper):
\begin{enumerate}
\item Because spacetime is nontrivially curved within the sphere, we cannot define the mass of the sphere simply as the integral of the local mass-density: indeed, that integral is not even well-defined in a coordinate-free way. Instead, the mass can defined by using Noether's theorem (according to which energy is the conserved quantity associated with time translation symmetry), calculated at a distance much larger than the shell radius at which the spacetime is approximately flat. The precise version of this concept of mass is called the \emph{ADM mass}, after \citeN{ADM} (a related version, the \emph{Bondi-Sachs mass} (\citeNP{bondi}, Sachs~\citeyearNP{sachs1961,sachs1962}), is better suited to handle situations involving radiation but rests on the same basic idea). If the sphere had non-trivial spatial momentum and/or angular momentum, analogous ADM momenta and angular momenta can also be defined, using the appropriate asymptotic Noether symmetries. 
\item The sphere becomes increasingly non-homogeneous, with the density being much higher towards the centre of the sphere. From this and the local equation of state (\ref{EMstate}), we can deduce that the locally-measured temperature also increases closer to the centre. The locally measured temperature $t(r)$  at a radius $r$ from the centre is related to the thermodynamic temperature (given by $1/T=\partial S/\partial U$) by
\be
t(r) = \alpha(m(r),r)^{-1} T
\ee
where $m(r)$ is the mass of the sphere internal to $r$ (more precisely: the ADM mass that the region of the sphere interior to $r$ would have if it were confined to that region and the rest of the sphere removed) and $\alpha(m,r)=(1-2Gm/r)^{1/2}$ is the gravitational redshift induced by a spherically symmetric mass $m$.
\item The sphere is no longer extensive in any meaningful sense: increasing $R$ to $KR$ and $M$ to $K^3M$ will not produce a qualitatively similar sphere. Indeed, if $R<\sim 0.254 R_S$, the sphere becomes unstable and undergoes gravitational collapse into a black hole.
\item The heat capacity of the sphere (\iec, the rate of change of mass with temperature at constant radius) decreases to zero and becomes negative, so that decreasing the energy of the sphere actually causes it to become hotter.
\end{enumerate}
Though Sorkin \emph{et al} do not discuss it, the notion of ``thermal contact'' also has to be analysed with some care for these systems. For a start, we cannot put two such spheres in thermal contact simply by placing them adjacent to one another: their mutual gravitation would radically alter each other's states, probably producing gravitational collapse unless handled carefully. An intermediate system is required. 

As a concrete example, consider the following process for transferring heat between two spheres with thermodynamic temperatures $T_1,T_2$, masses $M_1$, $M_2$ and surface redshifts $\alpha_1,\alpha_2$:
\begin{enumerate}
\item A box is slowly lowered to the surface of Sphere 1 from `infinity' (\iec, from very far above the sphere), allowed to fill with a small amount of radiation of local mass $m$ and temperature $ T_1 / \alpha_1$, and then slowly lifted back to infinity, requiring  \cite{unruhwald1982} work 
\be W_1 = (1-\alpha_1 )m.\ee
\item The box is adiabatically compressed or expanded (as appropriate) to a temperature $T_2/\alpha_2$, requiring additional (possibly negative) work 
\be
W_2 = \left( \frac{(T_2/\alpha_2)}{(T_1/\alpha_1)} - 1\right)m
\ee 
(as can be deduced from the equation of state (\ref{EMstate})) and changing its mass to $m (T_2/\alpha_2)(T_1/\alpha_1)$
\item The box is slowly lowered to the surface of Sphere 2, requiring negative work
\be
W_3 = - (1-\alpha_2 ) \frac{(T_2/\alpha_2)}{(T_1/\alpha_1)}m.
\ee
\item The box is then opened and the radiation released into Sphere 2; this is adiabatic, since it has the same local temperature as Sphere 2's surface.
\end{enumerate}
The entire process is adiabatic and has the following energy implications:
\be
\Delta M_1 = - \alpha_1 m; \,\,\,\,\ \Delta M_2 = \alpha_1 m (1 + (T_2/T_1)) ;\,\,\,\,  W = W_1+W_2+W_3 = \alpha_1 m (T_2/T_1). 
\ee
This has the characteristic form of a Carnot cycle. As a corollary, if $T_1>T_2$, net work is extracted by the process, and we can replace (3) by
\begin{enumerate}
\item[3'.] The box is slowly lowered towards the surface of Sphere 2 until the work extracted by doing so makes the whole process work-neutral, and then released to free-fall the rest of the way.
\end{enumerate}
The new process permits heat transfer, without work expenditure, from Sphere 1 to Sphere 2 provided $T_1>T_2$, and so provides a means to put the two spheres in (somewhat indirect) thermal contact.

In many examples of self-gravitating bodies, there is another way to put two bodies into thermal contact: seal them both into a very large box with reflecting walls, and wait. If one or other body is above absolute zero, it will emit electromagnetic radiation; in due course, the box will fill with radiation in local thermal equilibrium. Each body is in thermal contact with the radiation and so, indirectly, with the other body. This is an effective way (in principle and in thought, not in engineering practice!) to, for instance, place two neutron stars or white dwarfs into thermal contact. It is not really an option for our radiation spheres, because they are themselves comprised of thermal radiation so the breakdown into subsystems would not be well-defined.

\section{Classical black hole thermodynamics}\label{classical}

We can now consider whether, and to what extent, these thermodynamic notions apply to black holes and systems of black holes. In this section I consider only `classical' black holes, by which I mean: black holes, if we neglect or imagine away any quantum-field-theoretic effects: in particular, any matter fields present will be treated phenomenologically and classically. For clarity, I do not mean ``black holes, under the fiction that the world is exactly classical'': I'm not sure that is even well-defined (though see \citeN{curielclassical}) but in any case it presumably would not include thermal radiation, which can be treated phenomenologically as a classical fluid but whose derivation via statistical mechanics requires quantum theory.

\subsection{Black holes as objects}

The basic idea of BHT is that black holes are thermodynamic systems, and that a particular subclass of black holes (the stationary black holes) are the equilibrium states of those systems. But from the starting point of general relativity, it is hard to see how this is even coherent: in that context, a ``black hole'' is identified globally as a region of spacetime from which null geodesics cannot reach future infinity (see, \egc, \citeN{hawkingellis}). A spacetime region cannot itself change in time, so the notions of `equilibrium' or `equilibration' don't obviously make sense under this definition. 

But the relativist's concept of a black hole is not the only one extant in physics. \emph{Astrophysicists} have long spoken of black holes as objects which persist through time and whose properties change in time: any talk of black holes orbiting one another, or of two black holes merging to form a larger hole, or of the velocity of a black hole relative to another astrophysical object, seems to require a three-dimensional view of black holes as objects, in tension with the spacetime-region view natural in theoretical relativity.

The \emph{membrane paradigm} of Macdonald, Price and Thorne, developed in detail in the astrophysical context in \citeN{membraneparadigm} and adapted for the quantum theory of black holes by \citeN{susskindthorlaciusuglum}, addresses just this problem. Thorne \emph{et al} consider a timelike surface --- the `membrane', or `stretched horizon' --- that is placed around the true event horizon, at a very small proper distance from the true horizon. Thorne \emph{et al} give the stretched horizon an area $(1+\alpha)^2$ times that of the true horizon, where $\alpha$ is some positive real number $\ll 1$; more useful for foundational purposes is Susskind \emph{et al}'s convention (which I adopt henceforth), giving the horizon an area one Planck area larger than that of the true horizon.

The defining property of the event horizon, physically, is that nothing can emerge from it, and so in particular nothing can enter it and later return. But (in an admittedly somewhat heuristic sense), the stretched horizon is so close to the true horizon that \emph{virtually} nothing can cross the stretched horizon and return, because doing so would require extremely high accelerations (for timelike bodies dropped into the hole) or extremely short wavelengths (for photons, or extremely relativistic massive particles, on trajectories that pass between the stretched and true horizon) --- indeed, under Susskind \emph{et al}'s convention, it would require accelerations so high, and/or wavelengths so short, as to require Planck-scale physics to describe.\footnote{Thanks to Erik Curiel for pointing out the short-wavelength photon case.} So as long as we are dealing with energy levels well below the Planck scale, the stretched horizon may be treated as a one-way barrier just as can the true horizon.

On the other hand, the stretched horizon is an ordinary timelike surface; it can be treated as a two-dimensional closed surface in space that evolves through time, and so can be attributed potentially-time-dependent physical properties. And with its aid, we can then restate the goal of black hole thermodynamics as follows: to investigate the extent to which the stretched horizons of black holes can be treated as ordinary physical systems, and assigned mechanical, electromagnetic, and thermodynamic properties, from the point of view of any observers who remain outside the black hole --- or, to put it in less operational terms, the extent to which we can give a self-contained account of physics in the region of spacetime exterior to any black holes in terms of stretched horizons to which such properties are assigned.

\subsection{Equilibrium and equilibration for black holes}\label{blackhole-equilibrium}

Thermodynamics describes equilibrium systems in terms of their conserved quantities and external constraints. There are no real external constraints applicable to a black hole, but there are quantities which we would expect to be conserved: the energy, momentum and angular momentum of the hole (defined asymptotically by the ADM method) and its electrical charge. In each case these quantities are associated to long-range forces (gravity for the quantities associated to spacetime symmetries; electromagnetism for charge), as these forces ensure that matter bearing the conserved quantity will leave an asymptotic trace on the spacetime even once it crosses the stretched horizon. (Conserved quantities like baryon number, by contrast, cannot be expected to show up in the physics of the black hole exterior, since the long-range physics will be indifferent as to whether a particle that crosses the horizon is, say, a neutron rather than an anti-neutron.)
By working in a reference frame at which the black hole is at rest and its angular momentum is aligned along the $z$ axis (again, using the ADM charges to define this rigorously) we reduce the conserved quantities to three: the black hole's mass $M$, the magnitude $J$ of its angular momentum, and its charge $Q$. So if black holes have equilibrium states, we would expect the space of such states to be parametrised by these three quantities.

The definition of an `equilibrium' state is that it is unchanging in time, and general relativity offers a clear way to represent this: we look for \emph{stationary} solutions of the Einstein field equations, that is: solutions with a timelike Killing vector. Such solutions certainly exist for general $M,J,Q$: the \emph{Kerr-Newman} solutions to the coupled equations of general relativity and vacuum electromagnetism (aka Einstein-Maxwell theory) are stationary and parametrised precisely by mass, angular momentum and charge. When $Q=0$, these solutions reduce to the Kerr solutions of vacuum general relativity; when $J=0$, to the spherically-symmetric Reissner-Nordstrom solutions of the Einstein-Maxwell theory; when both are zero, to the well-known Schwarzschild solution. The Kerr-Newman solution only describes a black hole when $Q^2 + J^2/M^2 \leq M^2$, with solutions violating this inequality describing naked singularities; black holes that saturate the inequality are called \emph{extremal}, and are a somewhat puzzling  special case (one that has been of considerable importance in quantum gravity, as I discuss in the sequel to this paper).

The 1970s saw extensive work by Bardeen, Carter, Hawking, Israel and many others to prove the ``No-Hair Conjecture'': that the Kerr-Newman black holes are the unique stationary solutions to the Einstein-Maxwell theory, and so provide unique equilibria. To this day there remain loose ends in the conjecture and in its extension to more general situations in higher spacetime dimensions and with other long-range forces present, but in his review article in the Einstein Centenary Survey \cite{carter1979} felt able to say that
\begin{quote}
the no-hair theorems available \ldots are quite sufficient to justify --- with at least the degree of rigour usually considered acceptable in physics --- the assumption by any practically minded astrophysical theorist that any (external source free) black hole equilibrium-state solution \ldots belongs to the Kerr or Kerr-Newman families''.
\end{quote}
(See Carter's review article for detailed references and for a summary of the main results; see also \citeN{carter1997} for some historical remarks and \citeN{chrusciel} for a fairly up-to-date survey.)

Of course, \emph{thermodynamic} equilibrium requires more than mere stationarity: it requires \emph{non}-equilibrium systems to converge to equilibrium, and in particular, perturbations of equilibrium states to be damped back down to equilibrium. The stability of black holes, and the convergence to equilibrium of non-stationary black holes, has been extensively studied both analytically and numerically. By the mid-1980s (see chapters VI-VII of \citeN{membraneparadigm}, and references therein) it was established that perturbations of the stretched horizon by external gravitating bodies are damped away (for instance, the stretched horizon can oscillate, but these oscillations are damped, dying away back to equilibrium via the emission of gravity waves). Computer simulations of colliding black holes, and accretion of matter onto black holes, likewise demonstrate that the system evolves rapidly to the equilibrium-black-hole configuration, decaying by the emission of gravity waves (`ringdown'). And the historic observation of gravity waves in 2016 by the LIGO observatory \cite{ligo2016} provided a remarkably precise fit to the quantitative ringdown predictions, and so can reasonably be said to provide (ongoing) \emph{observational} support for black hole equilibration.

In summary: we have both a clear understanding of \emph{what} the black hole equilibria are, and a pretty good grasp on \emph{why} they are indeed equilibria: at the least, I think it would be hard to argue that we have any better theoretical control of how paradigm `normal' thermodynamical systems, like dilute gases, approach and remain at equilibrium. So far, black holes fully fit the requirements to count as thermodynamic systems.

\subsection{The laws of black hole thermodynamics}

To treat a black hole as a thermodynamic system requires us to identify external interventions, and to divide them into adiabatic changes and heat flows. The former is fairly straightforward: to move a black hole from one equilibrium state to another is going to require us to change its mass, angular momentum or charge, and the simplest way to do that is to drop matter into it. The latter is more delicate, since the division between `heat' and `work' is less obvious in an alien situation like this than for a box of gas. The simplest thing to do (in this case as in other less-familiar cases in `regular' thermodynamics) is to identify which transformations are reversible and which irreversible, and then define the quasi-static adiabatic processes as the reversible ones.

Christodolou and Ruffini demonstrated (\citeN{christodolouruffini}; see \citeN[pp.907--913]{mtw} for a discussion) that the quantity that plays the role of entropy for a black hole (at least for infinitesimal changes) is surface area (which, for an equilibrium black hole, is given by a known function of $M$, $J$ and $Q$): any intervention on an equilibrium black hole must leave the surface area nondecreasing, so that the reversible processes are those that leave surface area invariant and the irreversible processes strictly increase area. Reversible transformations of $J$ and $Q$ can be brought about as follows:
\begin{itemize}
\item To reversibly change the charge of a charged black hole, lower some charged matter very slowly on a cord so that it is suspended, stationary, just above the event horizon; then let go. 
\item To reversibly increase the angular momentum of a rotating black hole, fire some mass at it on a trajectory which just brushes the event horizon.
\item To reversibly decrease the angular momentum of a rotating black hole, use the \emph{Penrose process} (\citeN{penrose1969}, \citeN{penrosefloyd}; see \citeN[pp.267--271]{carrollgr} for an introduction): fall freely towards the black hole on a trajectory that passes just above the event horizon, and at point of closest approach, eject some mass into the black hole on a trajectory opposite to the direction of rotation of the hole.
\end{itemize}
Dropping charge into a black hole from finite height, or injecting mass on a non-brushing trajectory, or using the Penrose process on a higher trajectory, will in each case be \emph{irreversible}, bringing about an increase in surface area.

Hawking's area theorem \cite{hawkingarea} generalises Cristodolou and Ruffini's result beyond infinitesimal changes: Hawking proved that the area of any black hole is nondecreasing. His derivation presumes
\begin{enumerate}
\item that physics in the exterior of the black hole remains predictable (that is, roughly: assuming that no naked singularities form; see \citeN[pp.138-9]{waldqft} for a more precise discussion);
\item the \emph{null energy condition}: that the stress-energy tensor $T$ satisfies $T(v,v)\geq 0$ for any null $v$. This is violated in some exotic quantum-field-theoretic situations (of which more later) but seems a safe assumption for bulk matter, such as electromagnetic radiation and astrophysical fluids.
\item the Einstein field equations, which translate the null energy condition into a condition on the Einstein tensor. (Once that translation is made, the area theorem is purely a result in differential geometry, with no additional dynamical input.)
\end{enumerate}

 \citeN{bardeenlaws} christened the Area Theorem the ``Second Law of black hole thermodynamics''; in fact, it goes rather beyond the entropy-increase form of the standard Second Law, since black hole surface area remains well-defined even when a black hole is far from equilibrium, whereas thermodynamic entropy is defined only at equilibrium.

In the same paper, Bardeen \emph{et al} also established the ``First Law of black hole thermodynamics'' which states that
\be
\mathrm{d}M = \frac{1}{8\pi}\kappa \mathrm{d} A  - \Omega \mathrm{d}J - \Phi \mathrm{d}Q
\ee
where $\kappa$ is the surface gravity of the black hole, $A$ its surface area, $\Omega$ its angular velocity, and $\Phi$ the electric potential on its surface. This is \emph{precisely} the form of the standard First Law for a thermodynamic system where angular momentum and charge are conserved quantities, including the identification of the conjugates to $J$ and $Q$ as, respectively, angular velocity and electric potential. It permits us to identify the thermodynamic temperature of the hole as proportional to the surface gravity --- albeit, as long as we are considering a system in isolation, we have only identified entropy up to a monotonic function. Furthermore, we can independently prove the `physical-process' and `equilibrium-state' versions of the First Law distinguished by Wald (recall the discussion in section \ref{firstsecondlaw}), demonstrating that the overall structure of interventions on the black hole is self-consistent and fits the model of equilibrium thermodynamics. 

\subsection{Beyond Einstein's equation}

Bardeen \emph{et al}'s derivation of the laws of Black Hole thermodynamics presupposed the Einstein field equations; however, as Wald and collaborators have shown (\citeN{waldnoether}; see \citeN[pp.143-147]{waldqft} for an introduction and further references, and \citeN{jacobsonnoether} for more recent developments), the First Law (in both physical-process and equilibrium-state form) can be derived from a general diffeomorphism-invariant Lagrangian theory of gravity by identifying the entropy as (a form of) the Noether charge associated with the diffeomorphism symmetry, evaluated with respect to a vector field that coincides on the horizon with the horizon Killing vector.

 So far as I know there is no fully general non-decrease theorem for this generalised black hole entropy of the same scope of Hawking's area theorem, but \citeN{jacobsonhigherorder} have demonstrated that this generalised definition of entropy is nondecreasing under at least quasi-stationary processes, provided that the null energy condition is satisfied; they also prove the analog of Hawking's result for a large class of generalisations of the Einstein Lagrangian.

The physical reason for caring about this generalisation lies in the effective-field-theory program in contemporary particle physics. From that perspective, general relativity is thought of as a non-renormalisable effective field theory, regularised by a cutoff imposed by unknown Planck-level physics. In such a theory, all possible diffeomorphism-covariant action terms should be present; the Einstein-Hilbert action is just the leading-order term in an infinite expansion of the Lagrangian in these various terms. So the fact that black hole thermodynamics extends so naturally beyond the Einstein-Hilbert case is reassuring for the physical applicability of the theory. 

\subsection{Local properties of the stretched horizon}\label{localblackhole}

The stretched horizon of a black hole is, it seems, a purely \emph{fictional} entity, invisible to anyone falling through it and corresponding to no locally-present distribution of charge or energy. It is therefore frankly startling that it can be treated not simply as a formal device to make sense of black hole thermodynamics (as I used it above) but as an actual extended physical system with local thermodynamic properties.

To expand: as discussed \emph{in extenso} in \citeN{membraneparadigm} and references therein, we can treat the stretched horizon as a two-dimensional, electrically-conducting, viscous fluid, assigning to each infinitesimal part of its surface the exact charge, current, and stress-energy densities required to terminate the electromagnetic and gravitational field lines on its exterior. This assignment is arguably fictional since an observer freely falling through the horizon will not encounter these charges or energies, but from the point of view of physics outside the stretched horizon they are entirely real. To give some examples (many more can be found in Thorne \emph{et al}):
\begin{enumerate}
\item If a positively charged particle falls towards the North pole of an uncharged black hole, its field will induce a current flow of negative charge towards the north pole, which will become negatively charged; the South pole, opposite the direction of approach of the falling particle, will become positively charged. By applying the law of Ohmic dissipation to this current flow (the black hole's surface resistivity is $\sim$ 377 ohms) we deduce that heat will be dissipated in this process so that the black hole area increases. When the charged particle reaches the surface, current will flow back until the charge density on the surface is constant, dissipating more heat. Any region of charge excess will spread out exponentially so that the time for an initially non-equilibrium charge distribution to equilibrate is $\tau_{eq}=\sim M \log M$ in Planck units, or in more astrophysically useful units
\be\label{mlogm}
\tau_{eq} \sim 4.9 \times 10^{-6} \left(\frac{M}{M_\odot}\right)\left( \log (M/M_\odot) + 87.4\right) \mbox{seconds}
\ee
($M_\odot=1.99 \times 10^{30} \mathrm{kg}$ is the mass of the Sun). Only in the limit where the charge is lowered infinitely slowly to the surface will the current flow be so slow, and the readjustment of charge across the surface so complete, that no heat is dissipated; this is the reversible process described previously.(\citeNP{znajek1978,damour1978,macdonaldsuen1985}; \citeNP[pp.35--38,57--64]{membraneparadigm}.)
\item If an electrically neutral black hole rotates in an asymptotically constant magnetic field at right angles to its axis of rotation, eddy currents will be induced in the horizon. The magnetic field will exert a torque on the black hole via these currents, which will slow its rotation while also dissipating heat through electrical resistance. The result is that the rotational energy of the black hole will be dissipated as heat, slowing the black hole's rotation and increasing its area; the overall energy of the black hole remains conserved: that is, no energy is extracted from the static magnetic field in this process. (\citeNP{thornemacdonald1982},\cite[pp.102--106]{membraneparadigm}.)
\item If a black hole rotates in the tidal field of a larger gravitating body, the surface of the hole will be perturbed; this in turn produces viscous dissipation and corresponding viscous torque on the black hole in accord with the Navier-Stokes equation, dissipating heat and slowing the rotation of the hole. (\citeNP{hawkinghartle1972}; Hartle~\citeyearNP{hartle1973,hartle1974}; \citeNP[pp.252--255]{membraneparadigm}.)
\end{enumerate}
Also part of the local thermodynamics of black holes is the so-called \emph{Zeroth law of black hole thermodynamics} \cite{bardeenlaws}, which states that the temperature of a black hole is constant everywhere on the horizon. In ordinary thermodynamics, the analogous result --- that for a body at equilibrium, the local temperature is constant --- is more naturally thought of as a corollary of the Zeroth Law applied to the local-thermal-equilibrium context.
 
 \subsection{No thermal contact for classical black holes}\label{bekensteinbound}
 
 So far as we treat each black hole as an isolated system, the resemblance to a thermodynamic system seems pretty complete: black holes have notions of equilibrium and equilibration, reversibility and irreversibility, and local thermodynamic properties. But the resemblance terminates abruptly --- at least as far as classical black holes are concerned --- as soon as we try to consider them as thermodynamic systems interacting with other black holes, or with non-black-hole thermodynamic systems.
 
Specifically: there seems to be no available process that can reduce the entropy of one black hole and increase that of another (or of a non-black hole thermodynamic system), even if the total entropy is increasing. To the contrary, the analysis of reversible and irreversible processes above applied to each hole separately. Likewise, Hawking's area theorem applies separately to each connected component of a spacetime's event horizon, and so mandates not just that the total entropy of a system of black holes is nondecreasing but that the entropy of each black hole is separately nondecreasing. As a corollary, there seems no prospect of running a Carnot cycle between two black holes, and no prospect of allowing heat to flow from one hole to another. Likewise, there seems no way to make sense of heat flow from a black hole, to any other thermodynamic system. The nearest we can get is to allow two black holes to `interact' by colliding, in which case the area theorem guarantees that the new black hole has a larger entropy than its constituents, but this is a pale shadow of genuine thermal contact.
 
 In particular, classical black holes are completely black in the sense that they omit no thermal radiation. This means that a black hole placed in thermal contact with another body by the method of putting both in a box and letting it fill with radiation will simply eat all the radiation, however low its temperature. The only temperature that we seem consistently able to attribute to a classical black hole is then absolute zero.
 
These limitations are aggravated by Bekenstein's \citeyear{Bekenstein1973} observation that identifying black hole area with entropy also provides opportunities to violate the Second Law of thermodynamics unless we place some constraints on the form of the energy-entropy relation for ordinary matter --- constraints that do not seem well motivated within classical physics. Specifically:
 \begin{itemize}
 \item If some body of small mass $m$ and entropy $s$ is slowly lowered right to the event horizon and then released (the so-called `Geroch process', proposed by Robert Geroch during a 1970 Princeton colloquium), it will do work on the mechanism that lowers it. Qualitatively this is no different from the way in which a weight slowly lowered from a pulley can do work at the top of the pulley, but the quantitative scale is much larger: if a pointlike body of mass $m$ is slowly lowered to a point above the event horizon with redshift $\alpha$, then the work extracted is $W=m(1-\alpha)$ and so (by conservation of ADM mass) the mass increase of the black hole is $m\alpha$~\cite{unruhwald1982}. As the mass is lowered arbitrarily close to the horizon, $\alpha \rightarrow 0$, and so the black hole's mass after the process is carried out, and hence its surface area, will be unchanged --- but the entropy of the outside world will decrease by $s$. (This process can even be used to turn heat into work with perfect efficiency, thus violating at least the operational content of the Kelvin statement of the Second Law.)
 \item If some large body with mass $M$ and entropy $S$ undergoes gravitational collapse, it will form a black hole with area proportional to $M^2$, and decrease the entropy of the external world by $S$. If black hole area is identified with entropy (up to some scale factor $K$) then the total entropy change is $16 \pi KM^2 - S$, which for appropriate choices of $M$ and $S$ could easily be negative. \cite{susskindhologram}
 \end{itemize} 
 
As Bekenstein pointed out, both of these arguments would fail if there is some fundamental bound on the minimum size of a body with given entropy and mass. To expand: for simplicity let us specialise to a Schwarzschild (\iec, nonrotating, uncharged) black hole, where the metric is
\be \label{schwarzschildmetric}
\mathrm{d}s^2 = - \alpha(r)^2 \mathrm{d}t^2 + \alpha(r)^{-2} \mathrm{d}r^2 + r^2 (\mathrm{d}\theta^2 + \sin^2(\theta) \mathrm{d}\phi^2)
\ee
with $\alpha{r}=\sqrt{1-2GM/r}$. $\alpha(r)$ can be interpreted as the `redshift' at radial coordinate $r$, \iec the time dilation, relative to clocks at infinity, measured by an observer hovering above the black hole at constant radial coordinate $r$. The proper distance from the event horizon of an object at coordinate $r$ is
\be
d= \int_{2M}^r\frac{\mathrm{d}r}{\alpha(r)}.
\ee 
Very close to the black hole ($(r-2M)/2M \ll 1$), we can approximately take 
\be\alpha(r)\simeq\left( \frac{r-2M}{2M}\right)^{1/2},\ee evaluate $d$, and solve to get
\be
\alpha(d)=d/M.
\ee
So a spherical body of radius $d$, entropy $s$, and mass $m$, lowered slowly into the black hole, will increase the mass of the black hole by
$\delta M= md/M$, and so the black hole entropy by $\delta S_h = 8 \pi M \delta M= 8 \pi m d$. The total increase in (black hole entropy plus outside-matter entropy) is then
\be
\Delta S = \delta S_h - s =  8 \pi m d - s.
\ee
If some new principle of nature means that any such body must satisfy $s/m \leq 8\pi d$, that would suffice to ensure $\Delta S\geq 0$ (changing the geometry of the body changes the numerical coefficients but not the overall argument). A qualitatively similar constraint, $s/m \leq 2 \pi d$, also blocks Susskind's argument from gravitational collapse: the body, on forming a black hole, will have entropy $S_h=4\pi m^2$, so the net increase in entropy is
\be
\Delta S = 4\pi m^2 - s = 2 \pi m (2m -s/2\pi m) \geq 2\pi m (2m-d).
\ee
But the body must initially lie outside its own Schwarzschild radius, $d>2m$, to have avoided collapse already, so this must be positive. 

However suggestive this \emph{Bekenstein bound} might be, however, there is at least within classical physics no obvious reason why it must hold. And so to sum up: although classical black holes have some highly thermodynamic-\emph{like} properties, core aspects of thermodynamics depend on interactions between thermodynamic systems; these interactions do not seem to function correctly for classical black holes, rendering the analogy with thermodynamics purely formal.\footnote{\citeN{curielclassical} challenges this result and argues for a fully thermodynamic understanding of black holes even in the classical case; engagement with these arguments lies beyond the scope of this paper.}
 
\section{Quantum field theory}\label{quantum}

Quantum mechanics --- specifically, quantum field theory, formulated on a classical but curved spacetime --- removes the blemishes in BHT and transforms it from a suggestive analogy to a full equivalence. The central result here is the \emph{Hawking effect}: the discovery that black holes emit thermal radiation, at exactly the temperature that BHT would predict.
 
\subsection{Hawking radiation}

In this section I want to simply state what Hawking radiation is, and give some insight into its properties, leaving the question of whether we should believe it exists to the next section.
As a starting point to understand Hawking radiation, let's consider for simplicity a free, massless, scalar quantum field theory defined on Schwarzschild spacetime, with metric (\ref{schwarzschildmetric}). (Throughout this section, I assume a 'large' black hole, where curvatures outside the black hole are small compared to the Planck scale and hence quantum-gravitational effects can be neglected; the description of black hole radiation from Planck-scale black holes lies beyond currently-understood physics.)  The `external' region of that spacetime --- the region outside the event horizon, defined by $r>2GM$ --- is a globally hyperbolic spacetime suitable for describing the exterior of an uncharged non-rotating black hole. Since it has a timelike Killing vector --- corresponding to translation in the $t$ coordinate --- we can coherently analyse the eigenstates of energy of the theory, and since the field is free, those eigenstates can be defined by the occupation number of the various independent modes of the field, which are the definite-frequency solutions of the Klein-Gordon equation on the Schwarzschild background.

Given the linearity of the Klein-Gordon equation, and given the time-translation and rotational symmetries of Schwarzschild spacetime, any solution of the Klein-Gordon equation can be written (here I follow \citeN[pp.27-29]{harlowreview}) in the form
\be
\Psi(t,r,\theta,\phi)=\int \dr{\omega}\sum_{l,m} \alpha_{l,m}(\omega) f_{\omega l m}(t,r,\theta,\phi)
\ee
where
\be
f_{\omega l m}(t,r,\theta, \phi)= \frac{1}{r}Y_{lm}(\theta,\phi)\e{-i \omega t} \psi_{\omega l}(r)
\ee
and $Y_{lm}$ is a spherical harmonic.\footnote{For a reminder of the properties of spherical harmonics, see, \egc, \citeN{jackson}.} All of the detailed physics of the wave equation is contained in the functions $\psi_{\omega l}(r)$, with the rest following purely from the symmetry structure of the theory (recall that solutions to the Schr\"{o}dinger equation for a Coulomb potential, for instance, have the same form). So to understand the solutions, we need to understand the features of these functions.

To describe them further, it is helpful to introduce the \emph{tortoise coordinate} $r_*$, defined by
\be\label{tortoise}
r_* = r+ \ln |r/2GM - 1|,
\ee
which approximates $r$ for $r\gg 2GM$ but stretches the distance to the event horizon to cover the whole negative-$x$ axis; it also simplifies matters to adopt, temporarily, units in which $2GM=1$, \iec to use the Schwarzschild radius as our unit of distance. The radial function $\psi_{\omega l}$ then satisfies
\be\label{scatteringproblem}
\left( -\frac{\mathrm{d}^2}{\mathrm{d}r_*^2}+V(r)\right) \psi_{\omega l}  = \omega^2 \psi_{\omega l}  
\ee
where
\be\label{barrier}
V(r)=\frac{r-1}{r^3}\left(l(l+1) + \frac{1}{r}\right)
\ee
and where $r$ is given implicitly in terms of $r_*$ by (\ref{tortoise}). 

Formally (\ref{scatteringproblem}) is just the nonrelativistic Schr\"{o}dinger equation in one dimension, so that the problem of solving the Klein-Gordon equation has been reduced to a scattering problem in one dimension.  Modes can be thought of as incoming either from infinity or from the event horizon, and they will scatter off, or tunnel through, a potential barrier whose form depends on the angular momentum $l$. For $l\gg 1$ the barrier has height $\sim l^2$ and is located at $r=3/2$.  

We can now distinguish (following \cite[ch.VIII]{membraneparadigm}):
\begin{itemize}
\item IN modes, which come in from infinity and largely scatter off the angular-momentum barrier (for $l\gg 1$) with some small amplitude to penetrate the barrier and fall onto the event horizon;
\item UP modes, which come up from the vicinity of the event horizon and are largely trapped close to the horizon by the angular-momentum barrier (for $l\gg 1$) with some small amplitude to escape to infinity.
\end{itemize}
 
 Hawking's result is then the following: for a black hole formed by gravitational collapse, and with surface gravity $\kappa$, the quantum state of the exterior is a thermal state with respect to the UP modes, at a temperature $\kappa/2\pi$. (With respect to the IN modes, the quantum state is determined by boundary conditions; for an astrophysical black hole in the current epoch, for instance, we might take the IN modes to be in a thermal state at the temperature of the microwave background radiation.)
 
 To understand this summary, it is helpful to describe the radiation as seen by a fictional observer hovering at a fixed distance above the black hole. Such observers move along a trajectory of constant $r,\theta,\phi$, and are often called \emph{fiducial observers}, or FIDOs. A fiducial observer at a redshift of $\alpha$ follows an accelerated worldline with locally-measured acceleration $\alpha^{-1}\mathrm{d}\alpha/\mathrm{d}r$; we can imagine the observer being held in place by a rope supported at infinity. Fiducial observers observe very different effects depending on how close they are to the event horizon:
 \begin{itemize}
 \item A fiducial observer close to the event horizon (\iec, whose distance to the event horizon is small compared to the Schwarzschild radius, and in particular who is between the potential barrier described by equation (\ref{barrier}) and the event horizon) observes a thermal bath of black-body radiation which might be thought of as the black hole's ``atmosphere'': this radiation remains largely trapped by the potential barrier and mostly falls back into the black hole rather than escaping to infinity. The apparent temperature of the radiation, as measured by the fiducial observer, will be $T/\alpha= \kappa/2\pi \alpha$, because that observer's clocks are redshifted by a factor $\alpha$ compared to coordinate time; we have already seen that this shifting of the temperature is a general feature of self-gravitating thermal systems.
 
\item When the observer's redshift is large enough that the locally measured temperature approaches the Planck temperature, the field-theory model we have used becomes unreliable: put another way, at this redshift the locally-measured wavelength of the radiation approaches the Planck length and we expect quantum-gravitational effects to cut off the QFT description. This occurs (not by coincidence) when the fiducial observer has reached the \emph{stretched} horizon, using Susskind \emph{et al}'s convention for its location. 

\item Conversely, an observer \emph{far} from the event horizon sees a stream of outwardly flowing radiation appearing to emerge from the black hole. This radiation is \emph{not} black-body radiation, because modes of different angular momentum escape the black hole atmosphere to differing degrees. The \emph{grey-body factors} of a black hole describe how the black hole's emission spectrum, as a function of angular momentum and frequency, deviates from a perfect black body.
 \end{itemize}
 There is also a divergence between the observations of \emph{fiducial} observers, and those of \emph{inertial} observers falling into the black hole from far away, a divergence which increases as the event horizon is approached. In the outer (radiation) region of the spacetime, both groups of observers have similar experiences: they see an outward-going stream of radiation (although the increasing velocity of the infalling observer, and increasing acceleration of the fiducial observer, cause these experiences to diverge increasingly as they approach the black hole). Within the black hole atmosphere, and particularly as the observers approach the stretched horizon, the experiences become sharply different: while the fiducial observers experience ever-hotter thermal radiation, the infalling observer sees only slight  deviations from empty spacetime.
 
 \subsection{Evidence for the Hawking effect}
 
 Before considering the thermodynamics of black holes in the light of Hawking radiation, we should pause briefly to ask how confident we should be in its existence. After all, while the classical theory of black holes lies within the range of astrophysical observation and so is supported by quite a lot of direct evidence, there is no realistic prospect of observing Hawking radiation from astrophysical black holes, and so far no proposal for observing it in non-astrophysical contexts (\egc at the LHC, or through the decay of primordial black holes) has borne fruit. So the case is entirely theoretical; it is, nonetheless, very powerful.
 
 To my knowledge there are at least five independent, conceptually distinct routes by which the Hawking effect can be derived:
 \begin{enumerate}
 \item Hawking's original method of matching outgoing modes with exterior modes via the technique of Bogoliubov transformations (\citeN{hawking1975}; see \citeN[ch.7]{waldqft} for a review);
 \item Making precise the heuristic understanding of black hole evaporation by particles tunneling across the event horizon \cite{parikhwilczek};
 \item Requiring the quantum state of the black hole exterior to solve or nearly solve the semiclassical Einstein field equations, which is possible only if the outgoing modes are in a thermal state at the correct temperature (\citeN{candelas1980}, \citeN{sciamacandelasdeutsch}; see also section \ref{backreaction});
 \item Path-integral methods on the analytic continuation of the black hole exterior spacetime, which demonstrate that the radiation-free vacuum --- and, more generally, any thermal state at the wrong temperature --- leads to singularities at the horizon (\citeNP{hartlehawking1976,israel1976});
 \item Observing that radiation flow across the event horizon is necessary to prevent anomalous breaking of the diffeomorphism symmetry \cite{robinsonwilczek}. 
 \end{enumerate}
 Each has its strengths, weaknesses, and distinctive features. Hawking's original approach (1) is perhaps most directly tied to the physics of actual collapse-formed black holes, but is confined to free fields. At the other extreme, (4) is completely general but only applies to a black hole at thermal equilibrium with an external radiation bath, requiring additional physical justifications to be applied to collapse-formed black holes. (1) and (2) give concrete mechanisms for Hawking radiation, whereas (3)-(5) derive contradiction or unphysical paradox from its absence. But collectively, they strongly suggest that Hawking radiation really is a consequence of quantum field theory on curved spacetime, and not simply an artefact of a particular method of mathematical analysis. In turn, quantum field theory on curved spacetime is just an application of the general machinery of modern quantum field theory (in particular, the use of field theory to describe quantum fluctuations against a fixed classical background) and --- while it is fair to note that it has not passed the sort of precision tests which underpin support for, say, flat-space quantum electrodynamics, and that the notorious cosmological-constant problem\footnote{See~\citeN{weinbergcosmologicalconstant}, \citeN{PeeblesRatra}, and references therein; thanks to Erik Curiel for pointing out its significance here.} gives some grounds for concern about its overall coherence --- it is the theoretical underpinning for experimentally-tested results in astrophysics and cosmology, notably interferometry experiments involving photons that have passed through regions of curved spacetime.\footnote{Thanks to an anonymous referee for pressing the question of what grounds we have to accept curved-spacetime QFT.}
 
 It is also possible to give a fairly direct \emph{physical} argument for Hawking radiation. Consider a fiducial observer, very close to the event horizon (at some redshift $\alpha \ll 1$, say. The radius of curvature of the spacetime is much larger than the distance to the horizon, so locally it will appear to the observer as if they are accelerating in flat space at a constant locally-measured acceleration $\alpha^{-1}\mathrm{d}\alpha/\mathrm{d}r$. Sufficiently close to the event horizon, this tends to $\kappa/\alpha$, where $\kappa$ is the surface gravity. 
 The Unruh effect (\citeN{unruh1976}; see \citeN[pp.15--24]{harlowreview} for a helpful discussion, and \citeN{unruhreview} for an exhaustive review) tells us that an observer in flat spacetime with uniform acceleration $a$ experiences a bath of black-body thermal radiation at a temperature of $a/2 \pi$ (and the Unruh effect itself can also be derived in multiple ways: from Bogoliubov methods, via path integrals, and as a rigorous result in algebraic quantum field theory, to name three). So by the equivalence principle, we would expect our fiducial observer to see something very close to thermal radiation at this temperature: that is, at locally measured temperature $\kappa/2\pi \alpha$.
 
 Now, very close to the black hole the event horizon fills almost the whole sky, so we would expect most of the radiation observed by the fiducial observer to fall back into the black hole. But it doesn't \emph{quite} fill the whole sky, so any given radiation mode will have some amplitude to escape to infinity (with lower-angular-momentum modes having the highest amplitude). That radiation will be redshifted by a factor $\alpha$ and so will be seen at infinity to have a temperature $\kappa /2\pi$, in accordance with Hawking's prediction (and to be radially streaming from the black hole).
 
I pause to consider and rebut a well-known potential objection to the existence of Hawking radiation: the so-called \emph{trans-Planckian problem}. In a nutshell, the problem is that radiation observed from a black hole sufficiently long after it forms is apparently redshifted down from radiation at a locally-measured wavelength shorter than the Planck length, \iec a wavelengthl at which we should regard quantum field theory as unreliable in any case.  (And ``sufficiently long'' is not at all long, in astrophysical terms: the timescale is $\sim M \log M$ in Planck units, or (from equation (\ref{mlogm})) $\sim 10^{-3}$ seconds for solar-mass black holes.) At times much later than this, the original energy of the detected radiation gets \emph{bigger} than Planckian, indeed ridiculously big. 
 
If this argument were correct, it would demonstrate not simply that Hawking radiation is absent, but that there is some inherent inconsistency in defining quantum field theory on a curved background: as noted above, the \emph{absence} of Hawking radiation also leads to unphysical phenomena. But there are good reasons to doubt that it is correct. In particular (following \citeN{polchinskiniceslice}):
 \begin{enumerate}
 \item It is possible to foliate the spacetime of a collapse-formed black hole so that curvature and energy densities on each slice remain well-behaved and far from the Planck scale (at least for black holes that are themselves large compared to the Planck mass, and up to late stages in its evaporation, of which more later).
 \item The Hawking effect (if it exists) is low-energy physics, entirely describable in terms of the physics on each individual slice.
 \item So the form of the cutoff imposed on our quantum-field to regularise it at short wavelengths has no effect on the low-level physics, beyond the usual effect of rescaling the parameters of the field theory (which can be absorbed by renormalisation of those parameters).
 \item So it's harmless to use any cutoff we like, even the unphysical cutoff where we actually allow free-field theory to stay defined on arbitrarily short wavelengths. 
 \end{enumerate}
In a certain sense there is even \emph{empirical} evidence that the trans-Planckian problem is innocuous, and more generally that the arguments used to derive Hawking radiation are valid. Very close analogues of the Hawking effect occur in certain condensed-matter systems (as originally proposed by \citeN{unruh1981}) and have recently been empirically confirmed, even though in these theories it is unambiguous that the degrees of freedom are cut off at the atomic scale and that (the analogues of) trans-Planckian modes do not exist. (See \citeN{unruh2014} and \citeN{dardashtihawking}, and references therein, for more on these analogues and their conceptual significance.)

 For more on the trans-Planckian problem (and some residual worries) see \citeN[pp.46--54]{jacobsonqft}, \citeN[pp.37-39]{harlowreview}, and references therein; however, for the moment I think we are justified in setting it aside and regarding Hawking radiation as a nigh-unavoidable consequence of any attempt to do quantum field theory in the vicinity of a black hole event horizon. Physicists tend to regard the case for Hawking radiation as further bolstered by the unity it provides to black hole thermodynamics but even without that bolstering, the case is very strong --- though, of course, as good scientists we should remind ourselves that it remains purely theoretical, and that tests of quantum field theory itself in the curved-spacetime regime to date have been much less precise and numerous than in the flat-spacetime regime.
 
 \subsection{Back-reaction and evaporation}\label{backreaction}
 
Hawking's original calculation ---  and all the other calculations referenced above --- use quantum field theory on a fixed, non-dynamical background metric. As such, these derivations \emph{in of themselves} do not suffice to establish that Hawking radiation is fully analogous to ordinary thermal radiation, because they imply nothing about whether a radiating black hole ultimately decreases in mass and, thus, surface area. To establish this, we need to consider the back-reaction of the radiation on the metric field, and doing so in a fully satisfactory way requires a quantum theory of gravity, which of course we lack. Furthermore, given that there is no robust local definition of gravitational energy --- and, relatedly, no robust way to understand total energy as a sum of local energies --- we cannot \emph{simply} appeal to a local conservation law to conclude that radiating black holes evaporate.

Nonetheless we can give powerful arguments for that conclusion. The most direct is via appeal to Noether's theorem, applied on a sphere surrounding, and far from, the black hole: in that regime, we expect to be able to treat the hole as an approximately-isolated system in a larger region of Minkowski spacetime (see \citeN{wallacerpep} for more on this). So the symmetries of Minkowski spacetime allow us to write a global conservation law and to argue that the sum of the ADM mass-energy of the black hole plus the total energy of the radiation outside the sphere --- which is well defined, since that region is very nearly flat --- should be conserved, and hence that the energy flux through the sphere ought to equal the rate of decrease of the black hole mass.

We can make this more quantitative by considering the physics on the boundary of this large sphere (here, and for the rest of this section, for simplicity I confine my attention to uncharged, non-rotating black holes). In this regime, Hawking radiation just looks like a classical outflow of radiation, with stress-energy tensor
\be
T^{\mu\nu}=n^\mu n^\nu (A/r^2)
\ee
where $r$ is the Schwarzschild radial coordinate, $n^\mu$ is an outward-pointing null vector, and $A$ depends on the black hole mass (and lacks a simple analytic form, due to grey-body factors). The Schwarzschild metric does not solve the Einstein field equations with this stress-energy tensor, so the assumption that the black hole does not evaporate is inconsistent with classical general relativity in a regime where we expect the latter to hold. The unique spherically-symmetric solution to the field equations for this stress-energy tensor is the \emph{Vaidya metric} (see, \egc, \citeNP{joshiglobal}), which is basically the Schwarzschild metric with a time-dependent mass term $M(t)$ (`basically' because we need to express the metric in retarded coordinates, due to the finite speed of propagation of the radiation). And the time-dependence is given by
\be
\dbd{M}{t} = - 4\pi A,
\ee
exactly as would be predicted from a naive treatment of radiation as carrying away local mass-energy density.

To understand evaporation closer to the black hole, we need to go beyond the fully classical Einstein equation, as quantum-mechanical effects become relevant. The normal tool to investigate this is \emph{semiclassical gravity}, in which the classical metric is coupled by the Einstein field equations to the renormalised value of the quantum expectation value of the stress-energy tensor (possibly including first-order gravitational perturbations as an additional graviton field). That is, a solution of semiclassical gravity requires both a metric $g$ and a (Heisenberg) quantum state $\ket{\psi}$ such that
\be
G[g] = 8 \pi G \matel{\psi}{T[\op{\phi}]}{\psi}_{ren}
\ee
where $\op{\phi}$ schematically denotes the various quantum fields, $G$ is the Einstein tensor associated with $g$, and the `ren' subscript indicates that we need to renormalise the stress-energy tensor.

This theory can either be posited directly, on the plausible if heuristic grounds that quantum gravity `ought' to look like this when metric fluctuations are small, or derived as the leading non-classical term in certain expansion schemes for the effective quantum field theory of gravity coupled to matter (\citeNP{tomboulis1977,hartlehorowitz}); either way, it is the standard tool used for exploring back-reaction (see \citeN[ch.5]{waldqft} and references therein for detailed discussion). It is  difficult to calculate with; nonetheless, it has provided very strong evidence that radiating black holes do indeed radiate, and at exactly the rate predicted by the naive treatment.
In particular (and without pretending to be exhaustive):
\begin{enumerate}
\item Candelas, Deutsch and Sciama (\citeNP{candelas1980}; \citeNP{sciamacandelasdeutsch}) have calculated the stress-energy tensor for a scalar field on a Schwarzschild background near to the black hole event horizon. They find that the vacuum state of that field is strongly polarised, so as to have a very large negative stress-energy density, which diverges to negative infinity on the event horizon; this negative energy density is \emph{exactly} cancelled out by the positive stress-energy density of the quanta in a thermal state at the Hawking temperature. It follows from their results that 
\begin{enumerate}
\item the Hartle-Hawking state, in which both UP and IN modes of the field are in that thermal state, has zero net stress-energy density close to the black hole, and so solves the semiclassical equations;
\item any state which has any non-thermal UP mode (or any thermal UP mode at the wrong temperature) has divergent stress-energy density on the future horizon, and so fails to solve the semiclassical equations even approximately;
\item the Unruh state, in which the UP modes are thermally excited at the Hawking temperature but the IN modes are unexcited, has singular stress-energy density on the past horizon (which, for a collapse-formed black hole, is in any case unphysical) but only mildly nonzero stress-energy density on the future horizon, so that we would expect a self-consistent solution that is only a small perturbation of the Unruh state and the Schwarzschild solution;
\item In that small perturbation, the change in the area of the horizon can be calculated via the Newman-Penrose equation; the result is exactly in accord with the naive prediction from radiation flow. (This also gives insight into how the black hole's area can decrease in violation of the area theorem: the strongly polarised spacetime region close to the horizon allows a slight violation of the null energy condition.) 
\end{enumerate}
\citeN{frolovthorne} generalised these findings to rotating black holes.
\item Price, Thorne and Zurek (\citeNP{zurekthorne1985}; \citeNP[ch.VIII]{membraneparadigm}) translated this analysis into the membrane paradigm. In that framework, the positive stress-energy associated with the atmosphere of a black hole in the Hartle-Hawking state (\iec with both UP and IN modes thermal at the Hawking temperature) exactly cancels the negative stress-energy due to vacuum polarisation. If the IN states are unexcited, this results in a very slight depletion of the energy of the atmosphere and so a very slight negative energy flow across the stretched horizon. A precise set of conservation equations can be written at the stretched horizon that relate changes in its area to the flow of stress-energy across it; again, these reproduce exactly the naive prediction.
\item \citeN{abdolrahimi} use numerical methods to find the metric of a radiating black hole as a perturbation of the Schwarzschild metric; they obtain a metric which far from the event horizon becomes asymptotically close to the Vaidya metric, again with the expected rate of mass decrease.
\end{enumerate}
In conclusion: there are excellent reasons to think that the `naive' treatment of radiation gets the facts exactly right: black hole radiation carries away energy and decreases the mass and surface area of the radiating black hole.

\subsection{Hawking radiation and black hole thermodynamics}\label{blackhole-thermalcontact}

Hawking radiation slightly complicates the definition of `equilibrium' for black holes, but no more so than for any other radiating thermodynamic system. Any electromagnetically-interacting body above absolute zero will radiate, so if such a system is placed alone in the vacuum, it will eventually cool to absolute zero. We can handle this in three ways:
\begin{enumerate}
\item Place the system in a box (of arbitrary size) filled with thermal radiation at the same temperature as the system. The radiation and the system will be in thermal equilibrium with one another and the system will itself remain at equilibrium.
\item Place the system in an empty box that is not too large. It will fill up with thermal radiation at the same temperature as the system; if it is sufficiently small, this will happen without the system's temperature changing too much.
\item Finesse the issue by ignoring radiation, on the assumption that the timescale on which it cools the object is long compared to other timescales of interest.
\end{enumerate}
All these are available for black holes; the only subtlety is that the black hole's negative heat capacity means that it will be in \emph{unstable} equilibrium with a sufficiently large thermal bath. In normal circumstances, if a small fluctuation causes the radiating system to absorb a bit of heat, its temperature rises above that of the radiation bath, so it emits the heat back again; for a black hole, that fluctuation \emph{decreases} the temperature, so positive feedback will occur. However, if the box is sufficiently small, the decrease in temperature of the radiation bath exceeds that of the black hole and the system remains stable. Elementary calculations \cite{hawking1976} demonstrate that the total mass-energy of radiation in the box must be less than $1/4$ of the black hole mass; for a solar-mass black hole, the box must be no more than $\sim 10^{12}$ parsecs across, not an especially demanding constraint.

More importantly, Hawking radiation allows black holes to be in thermal contact with one another (and with other thermodynamic systems), in just the same ways as for other self-gravitating systems. The simplest way to do this is just to put the two systems (one or both of which is a black hole) in a large box, far enough from one another that their mutual gravitational interaction can be neglected.\footnote{This situation is a good illustration of my comment in the Introduction about mathematical rigor. In classical general relativity it is known \cite{mankoruiz} that there is no exactly-stationary vacuum solution describing two Kerr black holes (thanks to Erik Curiel for the reference). The general approach in (most of) the black-hole-thermodynamics literature is to dismiss this sort of concern on the grounds that (a) what is needed is not exact stationarity, but approximate stationarity, \iec negligible change of black hole orbit on the timescales relevant   to the problem at hand; (b) looking for exact solutions is in any case premature given that we do not have an exact theory of black holes which incorporates radiation and back-reaction. Since the decay timescale for binary black holes scales with the fifth power of their separation, but the time taken for a radiating black hole to equilibrate with its box scales with the cube of the box size, there does not seem to be any problem of principle in constructing a setup in which the binary system can indeed be treated as approximately stationary other than thermodynamic effects.} The box will fill up with radiation at a temperature intermediate between the two, and so heat will flow from the hotter body into the radiation and thence into the colder body. In particular, if the two bodies are at the same temperature, no energy will flow from one to the other: the Zeroth Law holds fully for black holes.

Alternately (and following Unruh and Wald~\citeyear{unruhwald1982,unruhwald1983}), we can achieve thermal contact via ``black hole mining'': slowly lowering a box on a rope into a black hole's atmosphere, letting it fill with thermal radiation, and slowly pulling it out again. The net energy extracted from the black hole in this process is easily calculated to be
\be
Q = \alpha (P + \rho) V
\ee
where $V$ is the box's volume and $\alpha, \rho$ and $P$ are respectively the redshift and the locally-measured radiation density and radiation pressure at the point where the box is removed. From the First Law of black hole thermodynamics, the change in the black hole's entropy is $Q/T_H$ (where $T_H$ is the hole's temperature); meanwhile, the box contains radiation at a temperature of $T_H/\alpha$. Thermal radiation has an entropy density $s=(P+\rho)/T$, so the entropy increase at infinity is also $Q/T_H$; in other words, this process is reversible, and indeed can be reversed just by slowly lowering a box of radiation into a black hole's atmosphere until its local temperature matches that of the atmosphere, opening it, and then slowly pulling out the empty box.

If instead we try to lower the box into \emph{another} black hole's atmosphere until we have extracted the same work as was required to lift the box in the first place, we will find that this is possible only if the second black hole is at a lower temperature than the first; if not, radiation pressure will support the box before we have extracted enough work. So --- just as with the radiation spheres --- this may be seen as a means of enabling heat flow from one black hole to another. 

Finally, if we lower the box into the second black hole until it is exactly supported by radiation pressure --- which is to say, until its temperature matches the local temperature of the atmosphere --- we will find that the net work done is 
\be
W = (\alpha_1 - \alpha_2)(P+\rho)V
\ee
where $\alpha_1$ and $\alpha_2$ are the redshifts at which the box is respectively filled and emptied. If the two black holes have temperature $T_1$, $T_2$, then we must have $T_1/\alpha_1=T_2/\alpha_2$, so that the heat $Q_1$ extracted from the first black hole, the heat transferred to the second black hole, and the work extracted satisfy
\be
Q_2 = (T_1/T_2) Q_1;\,\,\,\, W = (1-T_1/T_2) Q_1.
\ee
So this process is a Carnot process between the two black holes.

\subsection{The generalised second law and the Casini-Bekenstein bound}

At this point, we have established that stationary black holes behave \emph{almost} exactly like thermodynamic systems. But there is a loose end left over from section~\ref{bekensteinbound}: we have not yet established that the Second Law applies in full generality, nor seen how to block Geroch's and Susskind's thought-experiments which apparently allow violations of the Second Law. What would be needed to tie up this loose end would be a proof, in semiclassical gravity, of the ``generalised second law'': that the entropy of the black hole exterior plus the Bekenstein-Hawking entropy is non-decreasing. (As  \citeN[p.34]{harlowreview} notes, ``generalised second law'' is a bit misleading if black hole area really \emph{is} entropy, in which case this would just be the \emph{ordinary} second law. On the other hand, in semiclassical gravity it is the sum of a statistical-mechanical entropy with a purely phenomenological entropy, so it does have a hybrid nature.)

The three decades following Bekenstein's original conjecture saw a substantial if rather disunified literature on various thought experiments intended to support the generalised second law. For instance, Unruh and Wald \citeyear{unruhwald1982} argued that the Geroch process is prevented by Hawking radiation: the global entropy maximum of a (small) box of a given energy is achieved when the box is full of thermal radiation at that energy, and that box will float, supported by the radiation pressure of the black hole atmosphere, when it is deep enough into the atmosphere that its temperature matches the local atmosphere temperature. It can readily be shown that if the box is then opened so that its contents fall into the black hole, the entropy increase of the hole equals the entropy in the box. But this argument is controversial (see, \egc, \citeNP{bekenstein1999,marolfsorkin}) and in any case does not seem to address the case where a black hole is \emph{formed} by mass with a high entropy/energy ratio. Other authors offered various more-or-less rigorous arguments for the Bekenstein bound from quantum field theory, though for a long while Bekenstein's conjecture proved difficult to make precise, and at one point was thought to rely on some ceiling on the number of distinct fundamental particles (intuitively, the more particles there are, the more states there are at a given energy). See \citeN{walltenproofs} and references therein for a review of various attempts to prove the generalised second law over this period.

The last decade, however, has seen major progress in this area, largely due to increased insight into the way quantum entanglement changes with time in the black hole exterior. Building on work of \citeN{marolfminicross}, \citeN{casinibekenstein} was able to give a clear statement of the Bekenstein bound and then prove it fairly rigorously within quantum field theory. Similar ideas have been used by \citeN{wallgsl} to give a clear statement and fairly general proof of the generalised second law. 

While no doubt there is more to learn here, and plenty of interesting foundational work to do in understanding the recent results and their link to Bekenstein's work, there now seems to be pretty strong evidence that the generalised second law holds in semiclassical gravity in full generality, completing the case for a thermodynamic description of black holes.

\section{Conclusion}

Black hole thermodynamics is often described as a striking analog of ordinary thermodynamics. But if what it is to be a thermodynamic system is to obey the various laws of thermodynamics, and to interact with other thermodynamic systems in such a way that the combined system obeys those laws too, then stationary black holes are not \emph{analogous} to thermodynamic systems: they \emph{are} thermodynamic systems, in the fullest sense. More precisely, according to the best physics we currently have, a black hole at (or weakly perturbed from) equilibrium behaves \emph{exactly} like a conducting, viscous fluid at (or weakly perturbed from) equilibrium, arranged in a thin shell just outside the event horizon.

An obvious question follows. In all other cases we know, there is a statistical-mechanical underpinning both to the general laws of thermodynamics, and to the specific form of the equation of state and transport coefficients of each thermodynamic system. Can we likewise construct a black hole statistical mechanics to underpin black hole thermodynamics --- or are black holes fundamentally different from other thermodynamic systems at the microphysical level despite their common phenomenology? I address this topic in Part II.

\appendix
\section{Appendix: Dougherty and Callender on black hole thermodynamics}

In a thoughtful and provocative recent paper, Dougherty and Callender~(\citeyearNP{doughertycallender}; henceforth DC) reach the opposite conclusion to mine: that ``the analogy [between black hole thermodynamics and the ordinary kind] is not nearly as good as is commonly supposed.'' They advance three arguments: that BHT ``is often based on a kind of caricature of thermodynamics''; that it is ambiguous to what systems BHT is supposed to apply; that BHT is motivated by a controversial epistemic conception of entropy. Here I want to reply  to these arguments.

\subsection{A pale shadow of thermodynamics?}

DC point out many apparent weaknesses in the details of the analogy between black hole and ordinary thermodynamics, and it is simplest to respond to them in objection-reply form.
\begin{description}
\item[DC:] What is called the ``Zeroth law'' of BHT is analogous to a mere consequence of the real Zeroth Law.
\item[Response:] Fair enough (cf section \ref{localblackhole}). But the true Zeroth Law holds for black holes as much as for other thermodynamic systems, once Hawking radiation is allowed for (section \ref{blackhole-thermalcontact}).
\item[DC:] In ordinary thermodynamics, equilibrium systems minimise their internal energy; it's not clear whether that's even meaningful for black holes.
\item[Response:] Consider a black hole away from equilibrium. Assuming it eventually settles down to stationarity (for which, we have seen, there is strong evidence) then at late times the system will consist of outgoing gravitational radiation far from the black hole, plus a stationary black hole. To an arbitrarily good approximation we can then assign mass separately to the black hole and the radiation via Noether's theorem; gravitational radiation has positive energy, so the stationary black hole must have lower mass than its progenitor. This heuristic argument can be made precise via the Bondi mass (section \ref{td-gravity}): the \emph{Bondi mass loss formula} demonstrates that a system that emits gravitational waves has decreasing mass. See, \egc, \citeN{bondisachsreview} and references therein for a review of the techniques involved.
\item[DC:] There is no `in equilibrium with' relation for black holes.
\item[Response:] Hawking radiation lets us define such a relation in pretty much the same way it is defined for other gravitating and/or radiating bodies (section \ref{blackhole-thermalcontact}).
\item[DC:] In ordinary thermodynamics, internal energy is distinct from total energy; in BHT, it is identified with total energy.
\item[Response:] That's an artefact of working in the black hole rest frame, which is done purely for convenience (section \ref{blackhole-equilibrium}).
\item[DC:] If two black holes coalesce into one, the total entropy increases, even if the two black holes started off at the same temperature, `contrary to thermodynamics'.
\item[Response:] It's not contrary to thermodynamics. It's contrary to the thermodynamics of extensive systems, but black holes --- like self-gravitating systems in general --- aren't extensive (section \ref{td-gravity}).
\item[DC:] Substituting black hole entropy for area in thermodynamic laws makes a mess of thermodynamic relations where volume is a variable.
\item[Response:] Black hole entropy doesn't actually have the dimensions of area, unless we work in Planck units, in which case everything is dimensionless. But in any case, just because two quantities have the same units doesn't mean they can be substituted for one another in equations. (I confess I don't entirely understand DC's point here.)
\item[DC:] BHT is very non-extensive.
\item[Response:] Indeed it is, but (a) nothing in thermodynamics requires extensivity, and (b) extensivity fails in strongly self-gravitating systems for clear physical reasons (section \ref{td-gravity}), even before we consider black holes. (DC recognise this last point in a footnote, but claim that the subtleties of scaling in self-gravitating systems are `not analogous' to those of black holes. They don't say why; examples like the radiation sphere certainly look closely analogous.)
\end{description}
Beyond these specific points, if DC find that BHT is a caricature of ordinary thermodynamics, it is in part because the version of BHT they are discussing is itself a caricature, pretty much restricted to the laws of BHT stated in \citeN{bardeenlaws}. They don't consider Christodolou and Ruffini's discussion of reversible and irreversible processes, or any of the results of the membrane paradigm, or the various results on equilibration, and more importantly, while they note the existence of Hawking radiation they don't consider its role in thermal contact, or in permitting reversible heat flow to and from black holes via Unruh-Wald mining of the thermal atmosphere.

\subsection{Entropy of what?}

DC point out that the event horizon is a globally defined concept, that a concept like that is not a suitable basis for BHT, and that there is no generally-agreed-upon or unproblematic alternative. They are surely right to identify this as a profoundly important question with ramifications for our understanding of Hawking radiation, and perhaps for quanutm theory more generally. But it doesn't seem that relevant to black hole \emph{thermodynamics}. After all, thermodynamics is concerned with systems at equilibrium, and is essentially silent about non-equilibrium systems except to require that they go to equilibrium. So all BHT needs is a clear understanding of the horizon for stationary black holes (and, perhaps, for holes that are mildly perturbed away from equilibrium). But pretty much all candidate definitions for the horizon agree on stationary black holes. 

\subsection{Entropy and empiricism}

Bekenstein's original conjectures about black hole entropy made heavy use of the relation between information theory and entropy, and that link is frequently used as motivation in textbook discussions to this day. DC are critical both of information-theoretic approaches to entropy in general (they prefer a Boltzmannian conception of thermodynamics in which the information-entropy link is broken) and about its application to black holes in particular (they regard the idea of information being lost behind the event horizon as a particularly pernicious form of operationalism, given that we could just jump into the black hole ourselves at the same time that the "lost" information falls in). 

I think DC are being a little unfair here, both to Bekenstein himself (whose conjecture about black hole entropy was a consilience argument based on Christodolou, Ruffini and Hawking's results, and on various concrete thought-experiments, as much as on the general entropy-information link) and to operationalism (as shown by \citeN{haydenpreskill}, in the absence of Planck-scale effects, matter thrown into a black hole will be unobservable even by an observer who jumps in after it after only time $\sim M \log M$ ($\sim 10^{-3}$ second for astrophysical-scale black holes, recall); suggestively, this is the time it takes for the stretched horizon to equilibrate in the membrane paradigm, so information thrown into a black hole is lost \emph{in principle} after the black hole has equilibrated). 

But let's stipulate that they are entirely correct. That might be a reason not to have awarded a grant to Bekenstein (or Hawking) back in the 1970s. It doesn't seem a good argument against black hole thermodynamics now, after the discovery of the Hawking effect, the membrane paradigm and the Casini-Bekenstein bound. The case for black hole thermodynamics can now rest entirely on the concrete results that have been inspired by Bekenstein's conjecture, and does not need Bekenstein's original motivation. The history of science is full of ideas whose original motivation was shaky but which nonetheless worked out, and which now stand on their own without need for that original motivation.

To be fair to DC here, in their dialectic they take themselves already to have shown that the formal analogy between black hole thermodynamics and ordinary thermodynamics is weak, so that substantial additional motivation is needed to identify entropy with black hole area. So my criticisms of this section are not really independent of my earlier points.

\section*{Acknowledgements}

I am grateful to Harvey Brown, Craig Callender, Sean Carroll, Neil Dewar, Eleanor Knox, Carina Prunkl, Katie Robertson, and Chris Timpson for various useful discussions, and in particular to Jeremy Butterfield, Erik Curiel and an anonymous referee for many helpful suggestions.


\end{document}